# Enhancing diamond fluorescence via optimized nanorod dimer configurations

A. Szenes,[1] B. Bánhelyi,[2] T. Csendes,[2] G. Szabó,[1] and M. Csete[1,*]

[1]*Department of Optics and Quantum Electronics, University of Szeged, Dóm tér 9, Szeged, H-6720, Hungary*
[2]*Department of Computational Optimization, University of Szeged, Árpád tér 2, Szeged, H-6720, Hungary*
*[*]mcsete@physx.u-szeged.hu*

**Abstract**

Optical response of silicon (SiV) and nitrogen (NV) vacancy diamond color centers coupled to silver and gold nanorod dimers was numerically inspected. Optimization of the coupled emitter - nanorod dimer configurations was performed to attain the highest possible fluorescence enhancement by enhancing the excitation and emission of color centers simultaneously through plasmonic antenna resonances. To minimize losses conditional optimization was realized by setting a criterion regarding the minimum quantum efficiency of the coupled system (*cQE*). Restricted symmetric and allowed asymmetric antenna designs were also inspected to prove the potential advantages of asymmetric configurations tuneability. Among all inspected systems the highest $2.59\times10^8$ fluorescence enhancement with 46.08% *cQE* was achieved in case of NV color center coupled to asymmetric silver nanoantenna dimer. This is 3.17-times larger than the enhancement in corresponding symmetric configuration, which has larger 68.52% *cQE*. In case of SiV color center the highest $1.04\times10^8$ fluorescence enhancement with 37.83% *cQE* was achieved via asymmetric silver nanoantenna dimer. This is 1.06-times larger than the enhancement in the corresponding symmetric configuration, which has larger 57.46% *cQE*. The highest fluorescence enhancement achieved by gold nanorods is $4.75\times10^4$ with 21.8% *cQE*, which was shown in case of SiV color center coupled to asymmetric dimer. The attained enhancement is 8.48- (92.42-) times larger than the fluorescence enhancement achievable via symmetric (asymmetric) gold nanorod dimer coupled to SiV (NV) color center.

**Introduction**

Integrated photonic structures including single-photon emitters are crucial in fundamental research and in several application areas including magnetic sensing and quantum information processing (QIP) [1-4]. In QIP applications promising single-photon emitters are the nitrogen vacancy (NV) [5-7] and silicon vacancy (SiV) [8-11] diamond color centers due to their stable electron spin detectable even at room temperatures and due to the strong zero-phonon lines in the visible and in the near-infrared regions, respectively. The relatively narrow ZPL in case of SiV makes this color center particularly promising in specific applications. The light can resonantly couple into the collective oscillation of conductive electrons of metal nanoparticles, which phenomenon is the well-known localized surface plasmon resonance (LSPR) [12-14]. At the frequency of LSPR the E-field is significantly enhanced and confined into regions much smaller than the wavelength. Several examples in the literature prove that fluorescent light emission can be enhanced via plasmonic nanostructures [15-17]. Most straightforward approach is the parallel improvement of the excitation and emission [18]. The emitter may couple light into the LSPR of a nearby metal particle, which can reradiate this energy with improved efficiency and directivity due to its antenna properties at resonance, which is accompanied by volume charges accumulation [19-21]. The enhancement of the local density of optical states (LDOS) is much larger in the nanogap between two resonant metal nanoparticles, e.g. spheres, ellipsoids, rods and tips, than in proximity of a single nanoparticle [22-25]. In addition to this, the hybridization of surface plasmons of the individual particles can also lead to various coupled modes depending on interaction geometry, which promote larger fluorescence enhancement [26, 27]. The excitation rate of a diamond color center locating in this region can be considerably enhanced. The decay rate of the color center can be also modified in an inhomogeneous environment according to the Purcell effect [28-33]. Moreover, resolution enhancement in microscopy [34, 35] and spectral line-width narrowing can be achieved as well [36].

The importance of optimal coupling configuration has been already recognized in previous studies [24,33]. To fully exploit the capabilities of LSPR on metal nanoparticle dimers in diamond color center fluorescence improvement, adjoint geometry and illumination direction optimization is required. In this paper nanorod dimer configurations are numerically optimized to enhance the fluorescence of NV and SiV diamond color centers.

**Methods**

To perform efficient configuration optimization a numerical methodology was developed based on the commercially available COMSOL Multiphysics software. The complete optimization methodology and the implemented GLOBAL optimization algorithm is described elsewhere [32,37,38]. Briefly, the color centers were approximated as pure electric dipoles embedded into a diamond dielectric medium surrounding the two metal nanorods composed of two semi spheres connected by a cylinder. The optical response of a coupled system is primarily characterized via the emitter's *Purcell factor*, which is the ratio of dipole powers emitted in inhomogeneous environment ($P^{total}$) and in vacuum ($P_0^{radiative}$)

$$Purcell\ factor = \frac{P^{total}}{P_0^{radiative}} = \frac{P^{radiative} + P^{non-radiative}}{P_0^{radiative}}. \quad (1)$$

The coupled system's radiative rate enhancement ($\delta R$) is given as the power radiated into the far-field ($P^{radiative}$) divided by the dipole power emitted in vacuum:

$$\delta R = \frac{P^{radiative}}{P_0^{radiative}}, \quad (2)$$

while the coupled system is characterized by the quantum efficiency (*QE*), which is the ratio of *δR* and *Purcell factor*:

$$QE = \frac{P^{radiative}}{P^{radiative} + P^{non-radiative}}. \quad (3)$$

The values of *QE* were corrected with the intrinsic $QE_0$ of color centers during post-processing:

$$QE^{corrected} = \frac{P^{radiative}}{P^{radiative} + P^{non-radiative} + \frac{1-QE_0}{QE_0}}. \quad (4)$$

According to reciprocity the excitation can be treated as the emission, namely the analogous *Purcell*, *δR* and *QE* describe the system of a dipole emitting at the wavelength of excitation coupled to metal nanorod dimers.

The objective function of the optimization was the product of *δR*s at the excitation and emission, this $P_x$ *factor* describes the complete fluorescence enhancement. In many applications, high *QE* is essential hence conditional optimization was realized to achieve the highest $P_x$ *factor* by setting a criterion regarding the minimum *cQE* that has to be reached.

During the optimization, the nanorod long and short axis was tuned in [15 nm-160 nm] and in [13 nm-158 nm] interval, respectively, while the dimer gap was varied in [4 nm-20 nm] interval. In case of NV the dipole placed into the center of the gap between the two nanoparticles was parallel to the axis defined by the nanorods long axes. In contrast, in case of SiV two dipoles corresponding to excitation and emission were rotated by +/-45° to ensure balanced contribution to the $P_x$ *factor* according to their perpendicularity [9, 10].

First the dimer systems were restricted to be symmetric then the optimization was repeated for asymmetric dimers. The complete enhancement spectra between 500 and 800 nm were determined, then at the excitation and emission wavelengths the surface charge distribution (presented in left particle - gap - right particle sequence), near-field and far-field were also inspected to reveal the underlying nanophotonics. In the main text, we present the configurations corresponding to the highest achievable $P_x$ *factor* regardless the *cQE*. Data on the geometry and optical responses of these inspected systems (Tables S1-S3) as well as further details on configurations optimized with different *cQE* criteria (Figures S1-S4) are provided in Supplementary Material.

# Results

*SiV color center coupled to symmetric nanorod dimers*

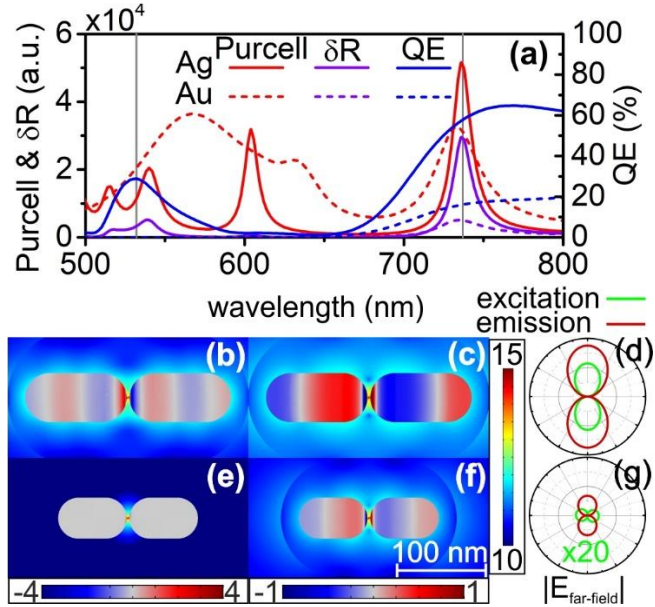

FIG. 1. Optical response of SiV color center and symmetric nanorod dimer coupled systems. (a) *Purcell factor*, radiative rate enhancement (*δR*) and quantum efficiency (*QE*) spectra. (b, c, e, f) Surface charge and near-field on logarithmic scale in arbitrary units, (d, g) far-field distribution; (b, e) at excitation and (c, f) at emission, (b, c, d) silver and (e, f, g) gold dimer (in (g) the signal at excitation is multiplied by 20 to improve visibility).

In case of the symmetric silver dimer and SiV color center coupled system the nanorods in the optimized configuration are elongated (Table S1). A local maximum on the *Purcell* (*QE*) spectrum is slightly detuned from the excitation, and the global maximum coincides with (detuned above) the emission (Fig. 1(a)). Both the excitation and emission are significantly enhanced, however the local maximum in *δR* is slightly more detuned from the excitation. The achieved $3.36 \times 10^3$ and $2.93 \times 10^4$ *δR*s with 7.2 nm and -3.6 nm detuning at the excitation and emission result in $9.83 \times 10^7$ $P_x$ factor with 57.46% *cQE*.

At the excitation on both nanorods $2 \times \lambda/2$ type volume charge distribution is accompanied by an enhanced parallel surface dipole, while at the emission the $\lambda/2$ antenna resonances are accompanied by a more commensurate parallel surface dipole (Fig. 1(b) and 1(c)). According to the dipolar far-field pattern perpendicular to the dimer axes the emitters are coupled both at the excitation and emission (Fig. 1(d)).

The same optimization of gold dimers resulted in a configuration consisting of slightly less elongated nanorods compared to the symmetric silver dimer (Table S1). The global maximum on the *Purcell* spectrum is above the excitation, while a local maximum coincides with the emission (Fig. 1(a)). The *QE* is low at the excitation, however it gradually increases by approaching the emission wavelength.

One single peak tuned to the emission is noticeable on the *δR* spectrum. The achieved 1.12 and 5.00×10$^3$ *δR*s with 72.2 nm and -3.6 nm detuning at the excitation and emission result in 5.60×10$^3$ *P$_x$ factor* with 16.01% *cQE*. At the excitation, the antenna does not show a coupled resonance, since only a local dipolar surface charge distribution is accumulated at the gap (Fig. 1(e)). Accordingly, weak near-field enhancement around the nanorods and dipolar far-field pattern along the dimer axis is observable (Fig. 1(g)). At the emission 1×λ/2 dipolar volume resonance appears on both composing nanorods, which is accompanied by a parallel local surface dipole (Fig. 1(f)). The coupling results in a dipolar far-field pattern perpendicularly to the long axes (Fig 1(g)).

The radiative rate enhancement is 1.76×10$^4$-times better in case of silver, 10.0$^3$-times smaller detuning is achievable at the excitation, while the same detuning is reached at the emission.

*SiV color center coupled to asymmetric nanorod dimers*

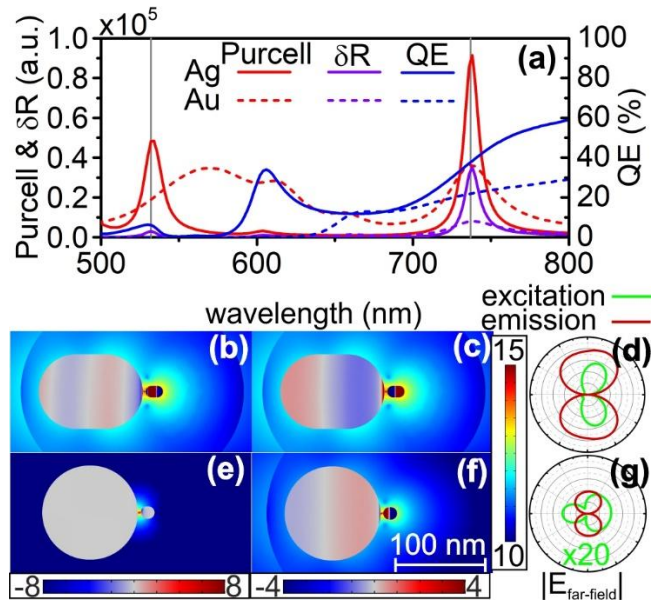

FIG. 2. Optical response of SiV color center and asymmetric nanorod dimer coupled systems. (a) *Purcell factor*, radiative rate enhancement (*δR*) and quantum efficiency (*QE*) spectra. (b, e, c, f) Surface charge and near-field on logarithmic scale in arbitrary units(d, g) far-field distribution; (b, e) at excitation and (c, f) at emission, (b, c, d) silver and (e, f, g) gold dimer (in (g) the signal at excitation is multiplied by 20 to improve visibility).

When asymmetric dimer configuration is allowed the procedure results in a strongly asymmetric silver nanorod geometry (Table S2). On the *Purcell* spectrum a local maximum is closer to the excitation compared to the symmetric configuration, while the emission almost coincides again with one of the resonance peaks (Fig. 2(a)). A local maximum in *QE* almost completely coincides with the excitation, while another one is far from the emission. The *δR* spectrum shows that the excitation and emission is significantly enhanced via a local and global maximum, respectively. The achieved $3.00\times10^3$ and $3.47\times10^4$ *δR*s with 0.0 nm and 0.8 nm detuning at the excitation and emission result in $1.04\times10^8$ $P_x$ *factor* with 37.83% *cQE*.

At the excitation, a charge distribution corresponding to $2\times\lambda/2$ ($1\times\lambda/2$) resonance appears on the larger (smaller) nanorod, which is locally enhanced by a parallel dipolar surface charge distribution at the gap (Fig. 2(b)). At the emission, a charge distribution corresponding to antiparallel $1\times\lambda/2$ resonance is observable on the nanorods with a reversing local charge accumulation at the gap (Fig. 2(c)). According to the coupled emitter-dimer system dipolar far-field pattern is observable at both wavelengths perpendicularly to the antenna axis with an asymmetrical scattering (Fig. 2(d)).

The optimization resulted in a strongly asymmetric configuration in case of gold nanorod dimers as well (Table S2). One local resonance is detuned from the excitation, while the global *Purcell* maximum almost coincides with the emission (Fig. 2(a)). No local *QE* maximum is observable at the excitation, while the emission is on the leg of a broad global *QE* maximum. On the radiative rate enhancement, a local maximum appears in between the excitation and emission, while the global maximum is tuned to the emission. The achieved 5.97 and $7.95\times10^3$ *δR*s with -5.2 nm and 0.6 nm detuning at the excitation and emission result in $4.75\times10^4$ $P_x$ *factor* with 21.80% *cQE*.

At the excitation, a weak dipolar surface charge accumulation appears at the gap, while no significant near-field enhancement arises on either of the nanorods, which indicates off-resonant configuration (Fig. 2(e)). The far-field pattern along the dimer axis is determined by the uncoupled emitter itself, and indicate asymmetrical scattering defined by the larger nanorod (Fig. 2(g)). At emission both nanorods are more strongly resonant. Furthermore, a reversal dipolar charge distribution develops on the coupled nanorods, while an inserted dipolar surface charge distribution parallel to that on the larger nanorod is also observable (Fig. 2(f)). The far-field emission pattern corresponds to a radiative dipolar coupled system, which emits throughout a wide polar angle region.

The radiative rate enhancement is $2.19\times10^3$-times better in case of silver, with amended and 1.33-times larger detuning at the excitation and at the emission, respectively.

*NV color center coupled to symmetric nanorod dimers*

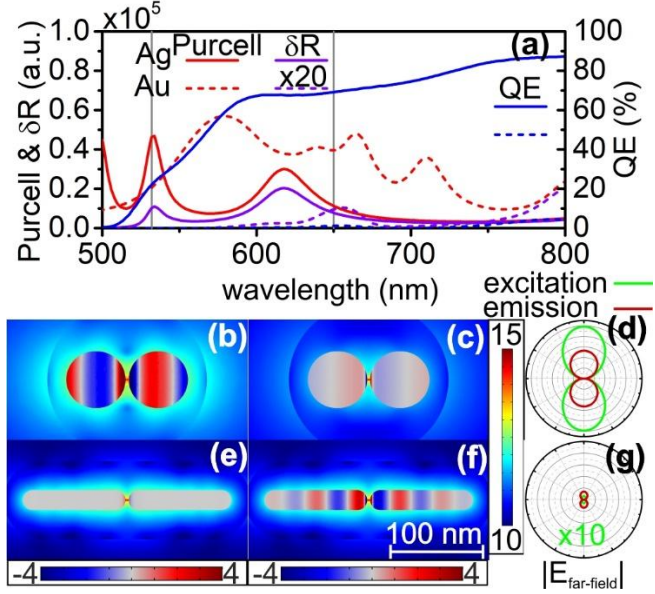

FIG. 3. Optical response of NV color center and symmetric nanorod dimer coupled systems. (a) *Purcell factor*, radiative rate enhancement (*δR*) and quantum efficiency (*QE*) spectra. (b, c, e, f) Surface charge and near-field on logarithmic scale in arbitrary units, (d, g) far-field distribution; (b, c) at excitation and (e, f) at emission, (b, c, d) silver and (e, f, g) gold nanorod dimer (in (a) and (g) Au and excitation signal is multiplied by 20 and 10 to improve visibility).

In case of a symmetric silver dimer and NV color center coupled system the optimized configuration consists of sphere-like nanorods (Table S1). At the excitation a narrow coincident *Purcell* resonance peak is observable, while the leg of a broader peak enhances the emission (Fig. 3(a)). The excitation is at the leg of a high *QE* resonance, while the emission is in between a local and a global *QE* maximum.

A local *δR* maximum is tuned to the excitation, while another local radiative rate peak is strongly detuned from the emission. The achieved $1.06 \times 10^4$ and $7.70 \times 10^3$ *δR*s with 1.8 nm and -32.2 nm detuning at the excitation and emission result in $8.19 \times 10^7$ $P_x$ *factor* with 68.52% *cQE*. The excitation is enhanced via a strong $1 \times \lambda/2$ antiparallel antenna resonances on the composing nanoparticles, which is accompanied by a strong and concentrated reversing surface dipole (Fig. 3(b)). The emission is enhanced via a weak $\lambda/2$ resonance on both composing nanorods, which is accompanied by a relatively stronger parallel surface dipole (Fig. 3(c)). Dipolar far-field patterns indicate uniquely stronger excitation enhancement, however the emission occurs into a relatively broader polar angle region (Fig. 3(d)).

The achieved $P_x$ *factor* is 0.83-times smaller, than in case of symmetric silver nanorod based dimer enhanced SiV.

The same optimization resulted in a configuration composed of strongly elongated nanorods in case of gold dimers (Table S1). The excitation is before the global resonance maximum, while the emission is in between two local resonance maxima, while a local *QE* maximum appears at the emission (Fig. 3(a)).

The emission is enhanced more significantly, however the global maximum on the radiative rate enhancement spectrum appears at a wavelength larger than the NV emission. The achieved 0.45 and $4.75 \times 10^2$ *δR*s with 83 nm and 6.8 nm detuning at the excitation and emission result in $2.14 \times 10^2$ *$P_x$ factor* with 1.19% *cQE*. The excitation is enhanced by a dipolar surface charge distribution at the gap of the composing nanorods (Fig. 3(e)). There is no significant near-field enhancement on either of the nanorods. According to the weakly coupled emitter-dimer the far-field signal is weak (Fig. 3(g)). The emission is enhanced via a $3 \times \lambda/2$ antenna resonance on both composing nanorods, one interfacial segment is strongly enhanced by the accumulated surface dipolar charge distribution at the gap (Fig. 3(f)). The resonance is accompanied by a strong near-field enhancement around the nanorods and by a dipolar far-field pattern perpendicular to the dimer axis, which is more intense than at excitation. The achieved *$P_x$ factor* is 0.04-times smaller, than in case of symmetric gold nanorod based dimer enhanced SiV. The radiative rate enhancement is $3.83 \times 10^5$-times better in case of silver, 46.11-times smaller detuning is achievable at the excitation, while 4.74-times larger detuning is available at the emission.

*NV color center coupled to symmetric nanorod dimers*

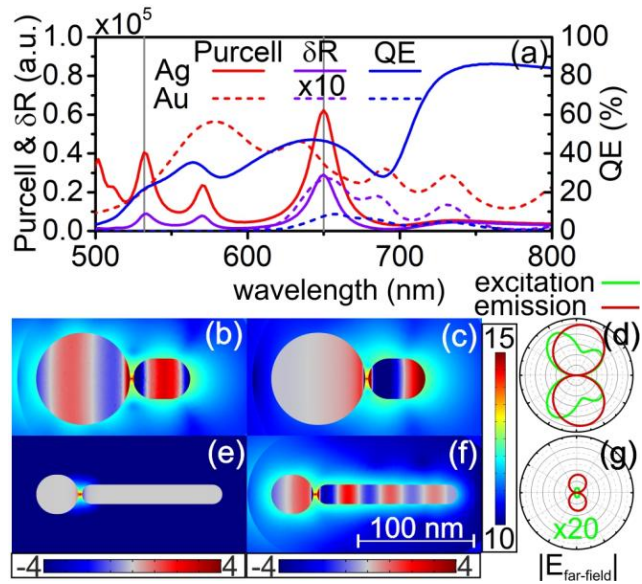

FIG. 4. Optical response of NV color center and asymmetric nanorod dimer coupled systems. (a) Purcell factor, radiative rate enhancement (*δR*) and quantum efficiency (*QE*) spectra. (b, c, e, f) Surface charge and near field on logarithmic scale in arbitrary units, (d, g) far-field distribution (b, e) at excitation and (c, f) at emission, (b, c, d) silver (e, f, g) gold nanorod dimer (in (a) and (g) signal of Au and excitation is multiplied by 10 and 20 to improve visibility).

The optimization of asymmetric silver dimer coupled to NV center resulted in a strongly asymmetric nanorod configuration (Table S2). Two *Purcell* resonance peaks are tuned to the excitation and emission wavelengths (Fig. 4(a)). The excitation is detuned from a local *QE* maximum, while the emission almost coincides with another one. The radiative rate at the excitation and emission is enhanced simultaneously. A local (the global) maximum coincides with the excitation (emission). The achieved $9.00 \times 10^3$ and $2.88 \times 10^4$ $\delta R$s with 1 nm and -0.4 nm detuning at the excitation and emission result in $2.59 \times 10^8$ $P_x$ *factor* with 46.08% *cQE*. The charge distribution at the excitation exhibits quadrupolar volume - parallel dipolar surface- dipolar volume modes accompanied by a quadrupolar far-field (Figs. 4(b) and 4(d)). The charge distribution at the emission exhibits dipolar volume - parallel dipolar surface- dipolar volume modes and the near-field shows that only the smaller nanorod is strongly resonant (Fig. 4(c)). The dipolar far-field pattern perpendicularly to the dimer axis indicates a coupled emitter-dimer system. The achieved $P_x$ *factor* is 2.72-times larger, than in case of asymmetric silver nanorod based dimer enhanced SiV.

The optimization resulted in a strongly asymmetric configuration in case of gold antenna dimer as well (Table S2). Neither the excitation nor the emission is coincident with a resonance (Fig. 4(a)). No local *QE* maximum appears at the excitation, however the global *QE* maximum is tuned to the emission.

Accordingly, no local peak appears on the $\delta R$ spectra below 650 nm and only the emission is enhanced significantly. The achieved 0.19 and $2.68 \times 10^3$ $\delta R$s with 118.8 nm and 1.8 nm detuning at the excitation and emission result in $5.14 \times 10^2$ $P_x$ *factor* with 7.66% *cQE*.

At the wavelength of excitation dipolar surface charge distribution is observable at the gap only (Fig. 4(e)). No significant near-field enhancement occurs, the far-field pattern is determined by the coupled emitter with a slightly asymmetrical scattering defined by the spherical nanoparticle (Fig. 4(g)). At the emission $1 \times \lambda/2$ dipolar volume - parallel dipolar surface - $3 \times \lambda/2$ volume modes are accompanied by a dipolar far-field pattern perpendicular to the dimer axis (Fig. 4(f)). The achieved $P_x$ *factor* is 0.01-times smaller, than in case of asymmetric gold nanorod based dimer enhanced SiV.

The radiative rate enhancement is $5.05 \times 10^5$-times better in case of silver, however 118.8- and 4.5-times larger detuning is achievable at the excitation and at the emission.

## Conclusions

In conclusion the constrained symmetric configuration of the optimized nanorod dimers predetermines the symmetry of the charge distribution, near-field enhancement and resistive heating as well. In the allowed asymmetric configurations the excitation and emission can be enhanced simultaneously more strongly in contrast to symmetric case due to the independently tunable nanorods. In all inspected coupled systems both the achievable $P_x$ *factor* and the *cQE* is higher in case of silver nanorod dimer antennas compared to gold antenna. Accordingly, the highest fluorescence enhancement in the inspected systems is $2.59 \times 10^8$ with 46.08% *cQE*, which is achievable via NV color center coupled to an asymmetric silver dimer antenna. The highest $1.08 \times 10^8$ SiV enhancement is achieved by an asymmetric silver nanorod dimer with 37.83% *cQE*. In case of gold nanorod dimers the highest $4.75 \times 10^4$ $P_x$ *factor* with 21.8% *cQE* is observable in SiV color center coupled to asymmetric design. Larger enhancement is achievable in case of SiV, except the asymmetric silver nanorod dimers. SiV color center coupled to gold nanorod dimers show one and two orders of magnitude larger fluorescence enhancement, than the NV coupled to gold nanorod dimers, in case of symmetric and asymmetric configurations.

## Acknowledgement

The research was supported by National Research, Development and Innovation Office-NKFIH through project "Optimized nanoplasmonics" K116362. Mária Csete acknowledges that the project was supported by the János Bolyai Research Scholarship of the Hungarian Academy of Sciences. The authors would like to thank Dávid Vass and Géza Veszprémi for figures preparation.

# Enhancing diamond fluorescence via optimized nanorod dimer configurations

András Szenes,[1] Balázs Bánhelyi,[2] Tibor Csendes,[2] Gábor Szabó,[1] Mária Csete[1,*]

[1] *Department of Optics and Quantum Electronics, University of Szeged, Dóm tér 9, Szeged, H-6720, Hungary*

[2] *Department of Computational Optimization, University of Szeged, Árpád tér 2, Szeged, H-6720, Hungary*

*mcsete@physx.u-szeged.hu

## *Supplementary Material*

### *SiV color center coupled to symmetric nanorod dimers*

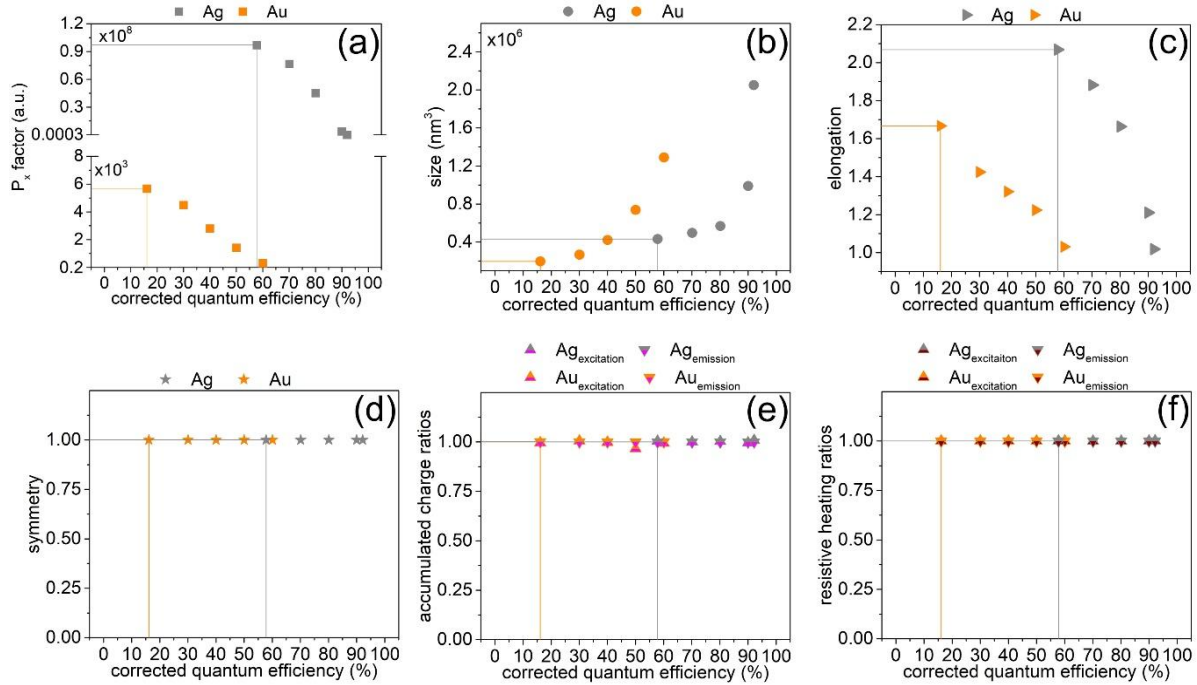

FIG. S1. Optical response and geometrical properties of systems consisting of SiV color center coupled to symmetric nanorod dimer optimized with different *cQE* criteria. (a) $P_x$ factor of the coupled systems, (b) size- and (c) elongation of individual nanorods. Ratio of (d) individual nanorods size, (e) accumulated charge and (f) resistive heating on the composing nanorods, each in smaller - to - larger nanorod sequence.

Silver is significantly better than gold to achieve higher values both in $P_x$ *factor* and in *cQE* via optimized nanorod dimers (Fig. S1(a)). The highest *cQE* achieved via optimized nanorod dimers is >92% in case of silver and >60% in case of gold. The $P_x$ *factor* almost linearly decreases as the *cQE* increases for both materials.

Larger particle size makes it possible to achieve larger *cQE* for both materials of gold and silver (Fig. S1(b)). Silver and gold nanorods composing the optimized dimers exhibit similar non-linear *size*(*cQE*) tendencies, but in different *cQE* intervals.

Less elongated i.e. more sphere-like particles are needed to achieve high *cQE* both in case of silver and gold nanorods (Fig. S1(c)). Silver is more elongated in the overlapping inspected *cQE* interval.

Both on the silver and gold nanorod based optimized symmetric dimer configurations the integrated surface charge density and resistive heating is the same on the individual nanorods (Fig. S1(d)-S1(f)).

In case of the optimized symmetric silver nanorod dimer the SiV emission phenomenon is accompanied by 3.83-times larger charge accumulation, while the resistive heating is 0.71-times smaller than at the excitation. In contrast, on the optimized symmetric gold nanorod dimer the SiV emission phenomenon is accompanied by 30.37-times larger charge accumulation on the average, while the resistive heating is 0.35-times smaller. These ratios are in accordance with that the contribution of the SiV excitation and emission enhancement is more balanced in case of the symmetric silver dimer, while the SiV fluorescence phenomenon is enhanced dominantly via emission improvement in case of the symmetric gold nanorod dimer.

## SiV color center coupled to asymmetric nanorod dimers

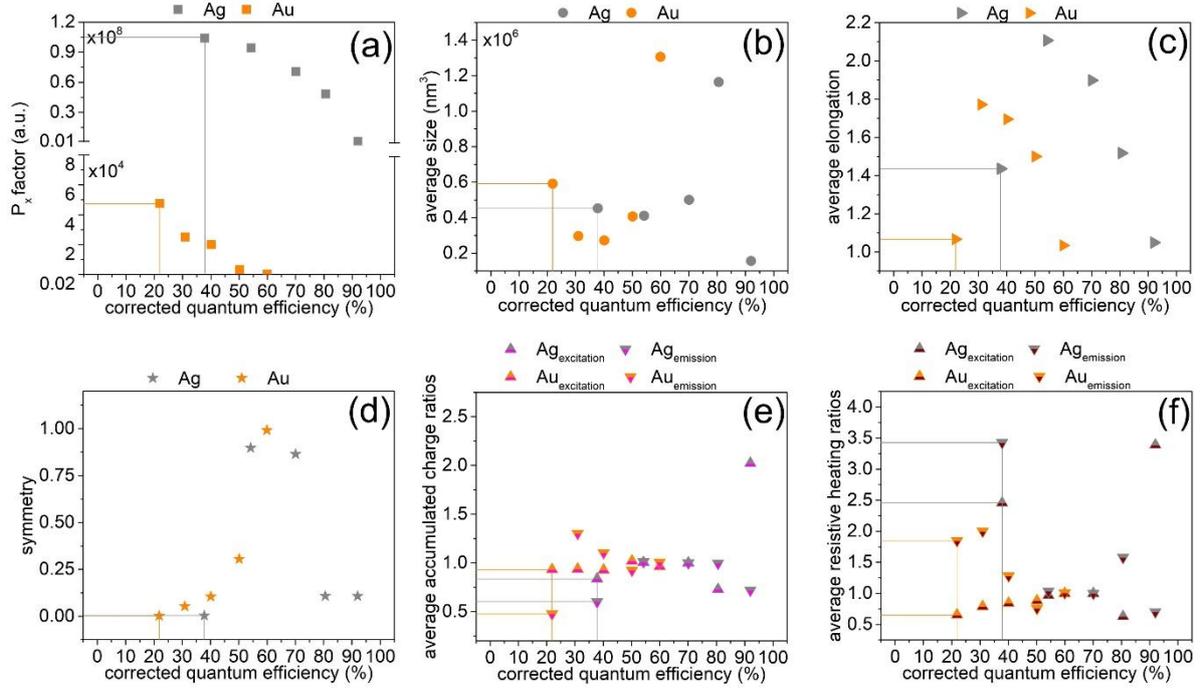

FIG. S2. Optical response and geometrical properties of systems consisting of SiV color center coupled to asymmetric nanorod dimer optimized with different *cQE* criteria. (a) $P_x$ factor of the coupled system, (b) average size and (c) average elongation of the nanorods. Ratio of (d) individual nanorod sizes, (e) accumulated charge and (f) resistive heating on the composing nanorods, each in smaller - to -larger nanorod sequence.

Silver is significantly better than gold to achieve higher values both in *$P_x$ factor* and *cQE* via optimized nanorod dimers (Fig. S2(a)). The highest achievable *cQE* in case of silver is >92%, while in case of gold it is >60%, which indicates that asymmetric silver dimer configuration makes it possible to attain slightly higher *cQE* values than the symmetric one. In addition to this, the highest *$P_x$ factor* is slightly larger in asymmetric configuration. The *$P_x$ factor* monotonically decreases by increasing *cQE* for both metals.

Similar tendency in size is observable for silver and gold nanorod dimers (Fig. S2(b)), however silver exhibits a maximum at intermediate *cQEs*, and interestingly the smallest average silver nanorod size corresponds to the highest achievable *cQE*. Gold exhibits a minimum in size at intermediate *cQE*, while the highest *cQE* is achievable via the largest average size of the gold nanorods.

Similar tendencies are observable in elongation (Fig. S2(c)), Namely, after a large maximum a monotonically decreasing elongation throughout the complete *cQE* interval is observable indicating that a gradually more sphere like rods are preferred. For the highest achievable *cQE* spherical nanoparticles are appropriate. In case of gold either to achieve the highest *$P_x$ factor* or the highest *cQE* spherical nanoparticles are needed.

Reversal tendencies are observable in symmetry (Fig. S2(d)). Wider symmetry interval allows to reach higher *cQE* in case of silver, while a gradually more symmetric configuration is preferred in case of gold.

In case of the optimized asymmetric silver nanorod dimer at the excitation the charge distribution is 0.84-times smaller on the smaller nanorod and the resistive heating is 2.46-times larger in it (Fig. S2(e)-S2(f)). Similarly at the emission the accumulated charge is 0.6-times smaller, while the resistive heating is 3.43-times larger in the smaller nanorod in accordance with the more absorptive nature of smaller nanoparticles. In case of the optimized asymmetric gold dimer at the excitation the accumulated charge is similar on the nanorods, it is only 0.93-times smaller, while the resistive heating is 0.65-times smaller on the smaller nanorod. At the emission the charge accumulation is 0.48-times weaker, conversely 1.85-times larger amount of heat is produced on the smaller nanorod, according to the strongly absorptive nature of the smaller metal nanoparticles (Fig. S2(e)-S2(f)).

In case of the optimized asymmetric silver nanorod dimer the SiV emission phenomenon is accompanied by 1.75- and 2.43-times larger charge accumulation, while the resistive heating is 0.37- and 0.26-times smaller than at the excitation. In contrast, on the optimized asymmetric gold nanorod dimer the SiV emission phenomenon is accompanied by 43.41- and 22.26-times larger charge accumulation, while the resistive heating is 0.24- and 0.67-times smaller. These ratios are in accordance with that the contribution of the SiV excitation and emission enhancement is much more balanced in case of asymmetric silver nanorod dimer, while the SiV fluorescence phenomenon is enhanced dominantly via emission improvement in case of the optimized asymmetric gold nanorod dimer.

## NV color center coupled to symmetric nanorod dimer

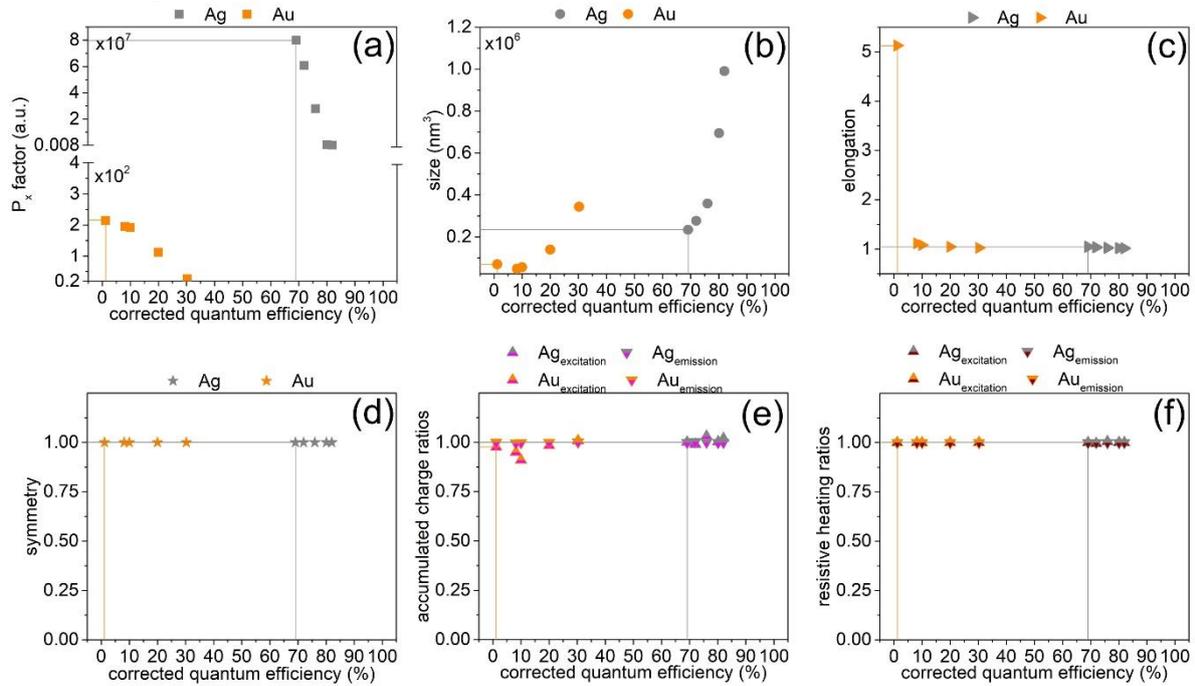

FIG. S3. Optical response and geometrical properties of systems consisting of NV color center coupled to symmetric nanorod dimer optimized with different *cQE* criteria. (a) $P_x$ factor of the coupled system, (b) size- and (c) elongation of individual nanorods. Ratio of (d) individual nanorod sizes, (e) accumulated charge and (f) resistive heating on the composing nanorods, each in smaller - to -larger nanorod sequence.

Also in case of NV coupling silver is significantly better to achieve higher values both in *$P_x$ factor* and in *cQE* via optimized nanorod dimers than gold (Fig. S3(a)). The highest achievable *cQE* is >80% and >30% in case of silver and gold, respectively. These values are smaller, while the difference in the maximum values is larger than in case of SiV color center - nanorod dimer coupled systems.

Larger nanorods correspond to larger *cQE* both for silver and gold dimers (Fig. S3(b)). Similar tendencies in *size*(*cQE*) are observable in symmetric silver and gold nanorod dimers, but in different intervals. Gold has a minimum at intermediate *cQE* value, while silver nanorods are typically larger in the optimized dimers.

Sphere geometry is preferred both in silver and gold nanorod based dimers (Fig S3(c)). One exception is the configuration of the gold nanorod dimer optimized for the highest achievable *$P_x$ factor*, which is strongly elongated.

Both on the silver and gold nanorod based optimized symmetric dimer configurations the integrated surface charge density and resistive heating is almost the same on the individual nanorods (Fig. S3(d)-S3(f)).

In case of the symmetric optimized silver nanorod dimer the NV emission phenomenon is accompanied by 0.16-times smaller charge accumulation, while the resistive heating is 0.04-times smaller than at the excitation. In contrast, on the optimized symmetric gold nanorod dimer the NV emission phenomenon is accompanied by 22.91-times larger charge accumulation, while the resistive heating is just 0.85-times smaller. These ratios indicate that uniquely the contribution of the NV excitation enhancement is more dominant in case of symmetric silver nanorod dimer, while the NV fluorescence phenomenon is enhanced exclusively via emission improvement in case of symmetric gold nanorod dimer.

## NV color center coupled to asymmetric nanorod dimers

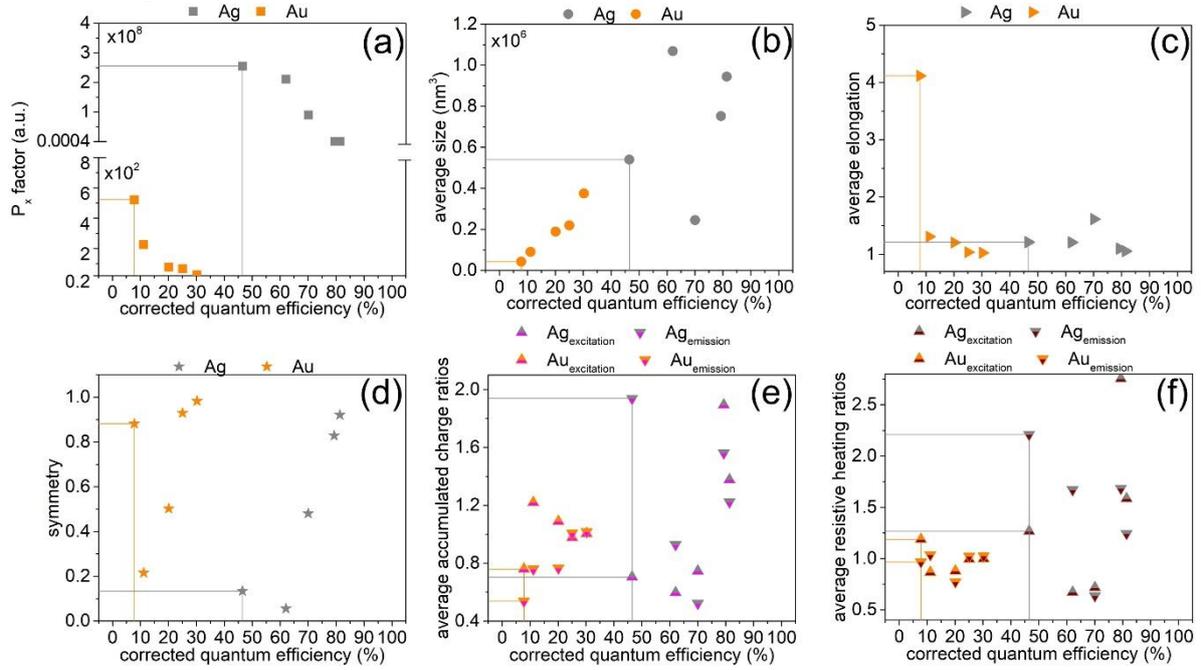

FIG. S4. Optical response and geometrical properties of systems consisting of NV color center coupled to asymmetric nanorod dimer optimized with different *cQE* criteria. (a) $P_x$ factor of the coupled system, (b) average size- and (c) average elongation of nanorods. Ratio of (d) smaller and larger nanorod sizes, (e) accumulated charge and (f) resistive heating on the composing nanorods, each in smaller to- larger nanorod sequence.

Also in case of NV coupling silver is significantly better than gold to achieve higher values both in $P_x$ *factor* and *cQE* via nanorod dimers (Fig. S4(a)). The highest achievable *cQE* in case of silver is around 80%, while in case of gold it is around 30%, indicating again that asymmetric nanorod dimer configuration makes it possible to attain larger $P_x$ *factor*. However it does not promote (allows slightly) higher *cQE*s in case of silver (gold). Larger difference is experienced in case of NV coupled asymmetric dimers with respect to corresponding symmetric configurations, than in case of SiV. As the *cQE* increases, the $P_x$ *factor* monotonically decreases for both materials

The preferred size interval is larger in case of silver nanorods, and larger average size corresponds to silver in the overlapping inspected *cQE* intervals (Fig. S4(b)). The average size linearly increases with *cQE* in case of gold.

Nearly sphere like nanorods are preferred in case of silver and gold as well. Only exception is a gold nanorod dimer configuration optimized for the highest (smallest) achievable $P_x$ *factor* (*cQE*) (Fig. S4(c)).

Similarly to the size, the *symmetry*(*cQE*) tendency is non monotonic in case of silver. In addition to this a larger symmetry interval allows to reach higher *cQE*. Asymmetric gold nanorod dimers with gradually increasing symmetry make it possible to achieve larger *cQE*. Only exception is the configuration optimized for the highest (smallest) achievable $P_x$ *factor* (*cQE*), which is almost completely symmetrical (Fig. S4(d)).

In case of the optimized asymmetric silver nanorod dimer at the excitation the accumulated charge is 0.71-times smaller on the smaller nanorod, in contrast the resistive heating is 1.27-times larger. At the emission 1.94-times more electric charge is accumulated on the small nanorod and approximately 2.21-times more heat generated on it. This is in accordance with that at the emission only the smaller nanorod is resonant. The larger resistive heating on the smaller nanorod at both wavelengths is in accordance with its smaller size (Fig. S4(e)-S4(f)). In case of the optimized asymmetric gold nanorod dimer 0.76-times less charge is accumulated, while 1.19-times more heat is absorbed on the smaller nanorod, similarly to silver. At the emission 0.54-times less electric charges are separated, however only 0.97-times less heat is absorbed on the smaller nanorod. This is in accordance with that at the emission only the larger nanorod is resonant (Fig. S4(e)-(f)).

In case of the optimized asymmetric silver nanorod dimer the NV emission phenomenon is accompanied by 0.54-times smaller and 1.48-times larger charge accumulation, while the resistive heating is 0.33 and 0.58-times smaller than at the excitation. In contrast, on the optimized asymmetric gold nanorod dimer the NV emission phenomenon is accompanied by 14.27 and 20.15-times larger charge accumulation, while the resistive heating is 0.62 and 0.76-times smaller. These ratios are in accordance with that the contribution of the NV excitation and emission enhancement is much more balanced in case of asymmetric silver nanorod dimers, while the NV fluorescence phenomenon is enhanced exclusively via emission improvement in case of the optimized asymmetric gold nanorod dimer.

| | | $P_x$ factor (a.u.) | | Purcell | QE (%) | δR (a.u.) | Δλ (nm) | nanorod1 \| $a_l$: 141.05, $a_s$: 68.20 nm | | | | nanorod2 \| $a_l$: 141.05, $a_s$: 68.2 nm | | | | d (nm) | symmetry | charge ratio | rh ratio |
|---|---|---|---|---|---|---|---|---|---|---|---|---|---|---|---|---|---|---|---|
| | | | | | | | | charge | | rh | | charge | | rh | | | | | |
| SiV | Ag | 9.83*10$^7$ | excitation: | 1.17*10$^4$ | 28.63 | 3355.16 | 7.20 | 6.31*10$^{-14}$ | charge em/ex | 9.39*10$^4$ | rh em/ex | 6.33*10$^{-14}$ | charge em/ex | 9.40*10$^4$ | rh em/ex | 2.00 | 1.00 | 1.00 | 1.00 |
| | | | emission: | 5.10*10$^4$ | 57.46 | 2.93*10$^4$ | -3.60 | 2.42*10$^{-13}$ | 3.83 | 6.66*10$^4$ | 0.71 | 2.42*10$^{-13}$ | 3.82 | 6.66*10$^4$ | 0.71 | | | 1.00 | 1.00 |
| | | | | | | | | elongation: | 2.07 | size: | 4.32*10$^5$ | elongation: | 2.07 | size: | 4.32*10$^5$ | | | | |
| | | $P_x$ factor (a.u.) | | Purcell | QE (%) | δR (a.u.) | Δλ (nm) | nanorod1 \| $a_l$: 95.28, $a_s$: 57.17 nm | | | | nanorod2 \| $a_l$: 95.28, $a_s$: 57.17 nm | | | | d (nm) | symmetry | charge ratio | rh ratio |
| | | | | | | | | charge | | rh | | charge | | rh | | | | | |
| | Au | 5596.60 | excitation: | 2.06*10$^4$ | 0.01 | 1.12 | 72.20 | 2.58*10$^{-15}$ | charge em/ex | 2.31*10$^5$ | rh em/ex | 2.57*10$^{-15}$ | charge em/ex | 2.31*10$^5$ | rh em/ex | 2.00 | 1.00 | 1.00 | 1.00 |
| | | | emission: | 3.12*10$^4$ | 16.01 | 4996.78 | -3.60 | 7.81*10$^{-14}$ | 30.29 | 7.98*10$^4$ | 0.35 | 7.81*10$^{-14}$ | 30.45 | 7.98*10$^4$ | 0.35 | | | 1.00 | 1.00 |
| | | | | | | | | elongation: | 1.67 | size: | 1.96*10$^5$ | elongation: | 1.67 | size: | 1.96*10$^5$ | | | | |
| | | $P_x$ factor (a.u.) | | Purcell | QE (%) | δR (a.u.) | Δλ (nm) | nanorod1 \| $a_l$: 78.29, $a_s$: 74.86 nm | | | | nanorod2 \| $a_l$: 78.29, $a_s$: 74.86 nm | | | | d (nm) | symmetry | charge ratio | rh ratio |
| | | | | | | | | charge | | rh | | charge | | rh | | | | | |
| NV | Ag | 8.19*10$^7$ | excitation: | 4.76*10$^4$ | 22.35 | 1.06*10$^4$ | 1.80 | 1.80*10$^{-13}$ | charge em/ex | 4.14*10$^5$ | rh em/ex | 1.80*10$^{-13}$ | charge em/ex | 4.14*10$^5$ | rh em/ex | 2.00 | 1.00 | 1.00 | 1.00 |
| | | | emission: | 1.12*10$^4$ | 68.52 | 7702.12 | -32.20 | 2.90*10$^{-14}$ | 0.16 | 1.80*10$^4$ | 0.04 | 2.90*10$^{-14}$ | 0.16 | 1.80*10$^4$ | 0.04 | | | 1.00 | 1.00 |
| | | | | | | | | elongation: | 1.05 | size: | 2.35*10$^5$ | elongation: | 1.05 | size: | 2.35*10$^5$ | | | | |
| | | $P_x$ factor (a.u.) | | Purcell | QE (%) | δR (a.u.) | Δλ (nm) | nanorod1 \| $a_l$: 135.87, $a_s$: 26.50 nm | | | | nanorod2 \| $a_l$: 135.87, $a_s$: 26.50 nm | | | | d (nm) | symmetry | charge ratio | rh ratio |
| | | | | | | | | charge | | rh | | charge | | rh | | | | | |
| | Au | 213.57 | excitation: | 2.09*10$^4$ | 2.16*10$^{-3}$ | 0.45 | 83.00 | 2.13*10$^{-15}$ | charge em/ex | 2.33*10$^5$ | rh em/ex | 2.08*10$^{-15}$ | charge em/ex | 2.33*10$^5$ | rh em/ex | 2.00 | 1.00 | 1.02 | 1.00 |
| | | | emission: | 3.98*10$^4$ | 1.19 | 474.51 | 6.80 | 4.83*10$^{-14}$ | 22.63 | 1.97*10$^5$ | 0.85 | 4.83*10$^{-14}$ | 23.18 | 1.97*10$^5$ | 0.85 | | | 1.00 | 1.00 |
| | | | | | | | | elongation: | 5.13 | size: | 7.01*10$^4$ | elongation: | 5.13 | size: | 7.01*10$^4$ | | | | |

TABLE. S1. Optical response and geometrical properties of systems consisting of SiV and NV color center coupled to symmetric nanorod dimer optimized with *cQE*=0 minimum criteria. $P_x$ factor – fluorescence enhancement, Purcell – Purcell factor, QE – quantum efficiency, δR – radiative rate enhancement, Δλ – detuning of resonance peak from excitation and emission, $a_l$ – long axis, $a_s$ – short axis, charge – accumulated charge on individual nanorod in arbitrary units, charge em/ex – accumulated charge ratio at emission/excitation on individual nanorod, rh – resistive heating on individual nanorod in arbitrary units, rh em/ex – resistive heating ratio at emission/excitaion on individual nanorods, d – distance of color center from metal, symmetry/charge ratio/rh ratio – ratio of size/accumulated charge/resistive heating each in smaller to-larger nanorod sequence.

| | | P$_x$ factor (a.u.) | | Purcell | QE (%) | δR (a.u.) | Δλ (nm) | nanorod1\| a$_l$: 23.79 , a$_s$: 16.11 nm | | | | nanorod2 \| a$_l$: 143.22 , a$_s$: 102.65 nm | | | | d (nm) | symmetry | charge ratio | rh ratio |
|---|---|---|---|---|---|---|---|---|---|---|---|---|---|---|---|---|---|---|---|
| | | | | | | | | charge | | rh | | charge | | rh | | | | | |
| SiV | Ag | 1.04*10$^8$ | excitation: | 4.91*10$^4$ | 6.11 | 2997.76 | 0.00 | 1.80*10$^{-14}$ | charge em/ex | 7.36*10$^5$ | rh em/ex | 2.16*10$^{-14}$ | charge em/ex | 3.00*10$^5$ | rh em/ex | 2.00 | 4.16*10$^{-3}$ | 0.84 | 2.46 |
| | | | emission: | 9.18*10$^4$ | 37.83 | 3.47*10$^4$ | 0.80 | 3.15*10$^{-14}$ | 1.75 | 2.72*10$^5$ | 0.37 | 5.23*10$^{-14}$ | 2.43 | 7.93*10$^4$ | 0.26 | | | 0.60 | 3.43 |
| | | | | | | | | elongation: | 1.48 | size: | 3.76*10$^3$ | elongation: | 1.40 | size: | 9.02*10$^5$ | | | | |
| | | P$_x$ factor (a.u.) | | Purcell | QE (%) | δR (a.u.) | Δλ (nm) | nanorod1\| a$_l$: 132.11 , a$_s$: 130.11 nm | | | | nanorod2 \| a$_l$: 19.21 , a$_s$: 17.18 nm | | | | d (nm) | symmetry | charge ratio | rh ratio |
| | | | | | | | | charge | | rh | | charge | | rh | | | | | |
| | Au | 5596.60 | excitation: | 1.91*10$^4$ | 0.03 | 5.97 | -5.20 | 2.11*10$^{-15}$ | charge em/ex | 2.59*10$^5$ | rh em/ex | 1.97*10$^{-15}$ | charge em/ex | 1.69*10$^5$ | rh em/ex | 2.00 | 2.65*10$^{-3}$ | 0.93 | 0.65 |
| | | | emission: | 3.65*10$^4$ | 21.80 | 7947.98 | 0.60 | 9.18*10$^{-14}$ | 43.41 | 6.10*10$^4$ | 0.24 | 4.39*10$^{-14}$ | 22.26 | 1.13*10$^5$ | 0.67 | | | 0.48 | 1.85 |
| | | | | | | | | elongation: | 1.02 | size: | 1.18*10$^6$ | elongation: | 1.12 | size: | 3.12*10$^3$ | | | | |
| | | P$_x$ factor (a.u.) | | Purcell | QE (%) | δR (a.u.) | Δλ (nm) | nanorod1 \| a$_l$: 123.13 , a$_s$: 121.05 nm | | | | nanorod2 \| a$_l$: 74.98 , a$_s$: 53.32 nm | | | | d (nm) | symmetry | charge ratio | rh ratio |
| | | | | | | | | charge | | rh | | charge | | rh | | | | | |
| NV | Ag | 8.19*10$^7$ | excitation: | 4.14*10$^4$ | 21.77 | 9001.79 | 1.00 | 2.20*10$^{-13}$ | charge em/ex | 3.20*10$^5$ | rh em/ex | 1.55*10$^{-13}$ | charge em/ex | 4.05*10$^5$ | rh em/ex | 2.00 | 0.13 | 0.71 | 1.27 |
| | | | emission: | 6.25*10$^4$ | 46.08 | 2.88*10$^4$ | -0.40 | 1.18*10$^{-13}$ | 0.54 | 1.06*10$^5$ | 0.33 | 2.29*10$^{-13}$ | 1.48 | 2.35*10$^5$ | 0.58 | | | 1.94 | 2.21 |
| | | | | | | | | elongation: | 1.02 | size: | 9.53*10$^5$ | elongation: | 1.41 | size: | 1.28*10$^5$ | | | | |
| | | P$_x$ factor (a.u.) | | Purcell | QE (%) | δR (a.u.) | Δλ (nm) | nanorod1 \| a$_l$: 43.82 , a$_s$: 41.45 nm | | | | nanorod2 \| a$_l$: 146.72 , a$_s$: 20.44 nm | | | | d (nm) | symmetry | charge ratio | rh ratio |
| | | | | | | | | charge | | rh | | charge | | rh | | | | | |
| | Au | 213.57 | excitation: | 2.13*10$^4$ | 9.02*10$^{-4}$ | 0.19 | 118.80 | 1.91*10$^{-15}$ | charge em/ex | 2.58*10$^5$ | rh em/ex | 2.51*10$^{-15}$ | charge em/ex | 2.17*10$^5$ | rh em/ex | 2.00 | 0.88 | 0.76 | 1.19 |
| | | | emission: | 3.50*10$^4$ | 7.66 | 2678.37 | 1.80 | 2.72*10$^{-14}$ | 14.27 | 1.59*10$^5$ | 0.62 | 5.06*10$^{-14}$ | 20.15 | 1.65*10$^5$ | 0.76 | | | 0.54 | 0.97 |
| | | | | | | | | elongation: | 1.06 | size: | 4.05*10$^4$ | elongation: | 7.18 | size: | 4.59*10$^4$ | | | | |

TABLE. S2. Optical response and geometrical properties of systems consisting of SiV and NV color center coupled to asymmetric nanorod dimer optimized with *cQE*=0 minimum criteria. *P*$_x$ factor – fluorescence enhancement, Purcell – Purcell factor, QE – quantum efficiency, δR – radiative rate enhancement, Δλ – detuning of resonance peak from excitation and emission, a$_l$ – long axis, a$_s$ – short axis, charge – accumulated charge on individual nanorod in arbitrary units, charge em/ex – accumulated charge ratio at emission/excitation on individual nanorod, rh – resistive heating on individual nanorod in arbitrary units, rh em/ex – resistive heating ratio at emission/excitaion on individual nanorods, d – distance of color center from metal, symmetry/charge ratio/rh ratio – ratio of size/accumulated charge/resistive heating each in smaller to-larger nanorod sequence.

| | | | P$_x$ factor | cQE emission | Δλ excitation | Δλ emission | | | | P$_x$ factor | cQE emission | Δλ excitation | Δλ emission | | | | P$_x$ factor | cQE emission | Δλ excitation | Δλ emission |
|---|---|---|---|---|---|---|---|---|---|---|---|---|---|---|---|---|---|---|---|---|
| Asym/Sym | SiV | Ag | 1.06 | 0.66 | 0.00 | -0.22 | Silver/Gold | SiV | Sym | 1.76*10$^4$ | 3.59 | 0.10 | 1.00 | NV/SiV | Sym | Ag | 0.83 | 1.19 | 0.25 | 8.94 |
| | | Au | 8.48 | 2.28*10$^3$ | -7.20*10$^{-2}$ | -0.17 | | | Asym | 2.19*10$^3$ | 1.74 | 0.00 | 1.33 | | | Au | 0.04 | 0.08 | 1.15 | -1.89 |
| | | | P$_x$ factor | cQE emission | Δλ excitation | Δλ emission | | | | P$_x$ factor | cQE emission | Δλ excitation | Δλ emission | | | | P$_x$ factor | cQE emission | Δλ excitation | Δλ emission |
| | NV | Ag | 3.17 | 0.67 | 0.56 | 1.24*10$^{-2}$ | | NV | Sym | 3.83*10$^5$ | 57.47 | 2.17*10$^{-2}$ | -4.74 | | Asym | Ag | 2.49 | 1.22 | - | -0.50 |
| | | Au | 2.41 | 6.43 | 1.43 | 0.26 | | | Asym | 5.05*10$^5$ | 6.01 | 8.42*10$^{-3}$ | -0.22 | | | Au | 0.01 | 0.35 | -22.85 | 3.00 |

TABLE. S3. Ratios of *P*$_x$ factors, corrected quantum efficiencies (*cQE*) and detunings (*Δλ*) of different configurations.